\documentclass[journal]{IEEEtran}
\usepackage[cmex10]{amsmath}
\usepackage{amsthm}
\usepackage{amssymb}
\usepackage{graphicx}
\usepackage{cite}
\usepackage{color}
\usepackage{algorithmic}
\usepackage{algorithm}

\usepackage{caption}

\begin{document}

\title{\LARGE{Sum Rate Maximization for Pinching Antennas Assisted RSMA System With Multiple Waveguides}}
\author{Peiyu Wang, Hong Wang, and Rongfang Song \vspace{-0.8cm}
\thanks{The authors are with the School of Communication and Information Engineering, Nanjing University of Posts and Telecommunications, Nanjing 210003, China (e-mail: wang\_py1999@163.com, wanghong@njupt.edu.cn, songrf@njupt.edu.cn).}
}

\maketitle

\begin{abstract}
In this letter, a pinching antennas (PAs) assisted rate splitting multiple access (RSMA) system with multiple waveguides is investigated to maximize sum rate. A two-step algorithm is proposed to determine PA activation scheme and optimize the waveguide beamforming. Specifically, a low complexity spatial correlation and distance based method is proposed for PA activation selection. After determining the PA activation status, a semi-definite programming (SDP) based successive convex approximation (SCA) is leveraged to obtain the optimal waveguide beamforming. Simulation results show that the proposed multiple waveguides based PAs assisted RSMA method achieves better performance than various benchmarking schemes.
\end{abstract}


\section{Introduction}
With the rapid increase in the number of communication devices, modern wireless systems face massive connectivity challenges. In terms of system capacity improvement, although space division multiple access (SDMA) can effectively utilize spatial resource to mitigate inter-user interference, the performance degrades significantly in overload scenarios. On the other hand, non-orthogonal multiple access (NOMA) demonstrates superior achievable rate compared to SDMA in overload scenarios. However, successive interference cancellation (SIC) technique, the key technology in NOMA systems, introduces substantial computational complexity in multiple users scenarios. To overcome the limitations of these multiple access technologies, rate splitting multiple access (RSMA) is proposed as an optimal trade-off between interference suppression and computational complexity\cite{mao2022rate}.

In order to validate the superiority of RSMA, \cite{lyu2024rate} and \cite{lyu2024rate2} investigated the performance of RSMA in both non-overloaded and overloaded scenarios, respectively. The results demonstrate that RSMA outperforms SDMA and NOMA in both overloaded and non-overloaded conditions. However, in RSMA systems, since the common stream needs to be decoded by all users, the achievable common rate is constrained by the user with the poorest channel condition. This limitation potentially reduces the system capacity. To mitigate the impact of poor channel users on the common rate, \cite{zheng2024joint} proposed a method based on successive convex approximation (SCA) and simulated annealing algorithm to selectively apply RSMA to partial users with the remaining users employing the conventional multiple access scheme. Furthermore, a novel cooperative RSMA scheme is proposed in \cite{abbasi2022transmission}, in which users experiencing better channel conditions serve as relays, assisting BS signal transmission to users with poorer channel quality.

In RSMA systems, since the decoding process of private streams can be regarded as a realization of SDMA, the private rate can achieve better performance under low spatial correlation conditions. Although fluid antenna (FA) and movable antenna (MA) can dynamically change channel states, the adjustment is limited in a small deployment range \cite{liu2025pinching}. Furthermore, due to the limited adjustment range, FA and MA cannot significantly improve channel strength, thereby constraining the common rate performance in RSMA systems. Recently, researchers proposed a novel pinching antenna (PA) technology that establishes line-of-sight (LoS) communication links with users by coupling radio-frequency signals transmitted through waveguides into free space, which significantly enhances system deployment flexibility \cite{ding2025flexible}. A waveguide division multiple access (WDMA) scheme was proposed in \cite{zhao2025waveguide}, where continuous PA placement scheme and discrete PA activation strategy were investigated to maximize the sum rate. Although the proposed scheme effectively mitigates interference, it is not suitable for overload scenarios. Different from \cite{zhao2025waveguide}, \cite{wang2025antenna} proposed a PAs assisted NOMA system with single waveguide, where the more practical discrete PA configuration is considered to maximize sum rate.

In this letter, we consided a PAs aided RSMA system, in which the private rate can be improved by optimizing PA activation scheme to reduce channel correlation. For the common rate, the PAs can also be leveraged to improve the channel conditions of users with poor channel quality by reducing the transmission distance, thereby improving the rate performance. Motivated by these, a PAs assisted RSMA system with multiple waveguides is proposed for the first time to maximize sum rate. The main contributions of this letter are as follows: i) A novel PAs aided RSMA system with multiple waveguides is proposed to maximize the sum rate by optimizing PA activation scheme and waveguides beamforming. ii) A more practical discrete PA system is considered. Moreover, in order to address the issue of exhaustive solutions which are infeasible due to high computational complexity, a low-complexity PA activation scheme based on spatial correlation and distance is proposed. iii) In overload scenarios, the conventional beamforming schemes become inadequate, thus an SCA based semi-definite programming (SDP) algorithm is proposed to optimal waveguide beamforming. Simulation results show that compared with various baselines, the PAs aided RSMA systems with multiple waveguides can achieve significant rate improvement.

\section{System Model}

\begin{figure}[!tbp]
  \centering
  \includegraphics[width=0.35\textwidth]{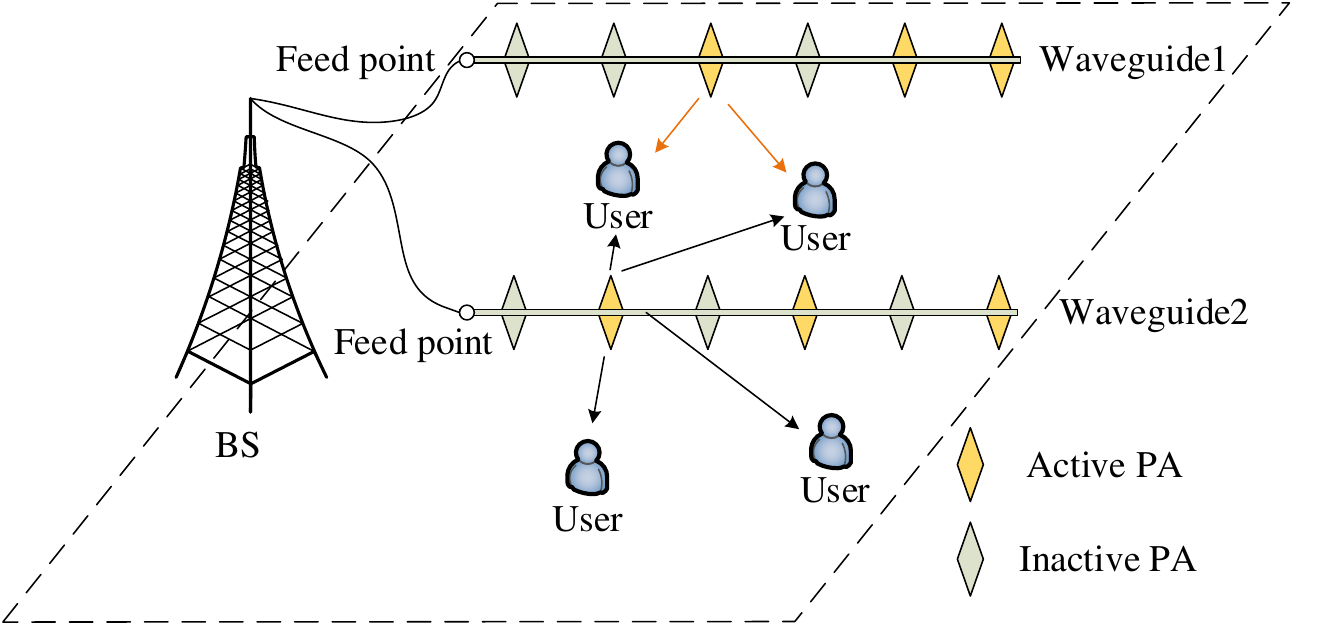}
  \vspace{-0.2cm}
  \caption{Illustration of downlink transmission for pinching antennas aided RSMA system.}\label{WANG_fig1}\vspace{-0.4cm}
\end{figure}

In this letter, the PAs assisted RSMA downlink systems with multiple waveguides are considered, which are composed of a base station (BS) and $K$ users. All users are equipped with a single antenna. The BS transmits signal with the assistance of $M$ waveguides, where $K > M$, and $N$ to-be-activated PAs are available on the each waveguide, as shown in Fig. \ref{WANG_fig1}. The location of the $n$-th pinching antenna on the $m$-th waveguide and the feed point (FP) of the $m$-th waveguide are ${\chi}^{PA}_{n,m}=(x_{n,m}^{PA},y_{n,m}^{PA},z_{n,m}^{PA})$ and ${\chi}^{FP}_{m}=(x_{m}^{FP},y_{m}^{FP},z_{m}^{FP})$, respectively. Users are uniformly distributed in a square region whose side length equals the waveguide length $D$, and the location of the $k$-th user is ${\chi}_{k}=(x_{k},y_{k},0)$. Furthermore, the PAs are uniformly distributed along each waveguide and the minimum spacing between any two PAs is $\Delta\geq \frac{\lambda}{2}$, where $\lambda$ is wavelength.

\subsection{Channel Models}

The channel model from the $n$-th PA on the $m$-th waveguide to the $k$-th user can be expressed as
\begin{align}\label{eq1}
{h}_{n,m\to k}=\frac{\lambda}{4\pi}\cdot\frac{ e^{-j\frac{2\pi}{\lambda}|{\chi}^{PA}_{n,m}-{\chi}_{k}|}}{|{\chi}^{PA}_{n,m}-{\chi}_{k}|},
\end{align}
where $\lambda=\frac{c}{f_c}$ is the wavelength, $f_c$ is the carrier frequency, and $c$ is the speed of the light \cite{xu2025rate}. $|{\chi}^{PA}_{n,m}-{\chi}_{k}|=\sqrt{(x_{n,m}^{PA}-x_k)^2+(y_{n,m}^{PA}-y_k)^2+(z_{n,m}^{PA})^2}$ is the distance between the $n$-th PA on the $m$-th waveguide and the $k$-th user.

Let $g_{n,m}$ denote the propagation channel from the feed point to the $n$-th PA on the $m$-th waveguide, which can be expressed as
\begin{align}\label{eq1}
{g}_{n,m}=e^{-j2\pi \frac{\eta_{eff}|\chi^{FP}_{m}-\chi^{PA}_{n,m}|}{\lambda}},
\end{align}
where $\eta_{eff} > 1$ is the effective refractive index of the waveguide \cite{wang2025antenna}, $|{\chi}^{FP}_{m}-{\chi}^{PA}_{n,m}|=\sqrt{({x}^{FP}_{m}-{x}^{PA}_{n,m})^2+(y^{FP}_{m}-{y}^{PA}_{n,m})^2+({z}^{FP}_{m}-{z}^{PA}_{n,m})^2}$ is the distance between the FP on the $m$-th waveguide and the $n$-th PA on the $m$-th waveguide.

Thus, the channel of the $k$-th user can be given as
\begin{align}\label{channel}
\mathbf{h}_k=( \begin{smallmatrix}
\sum_{n=1}^{N}\delta_{n,1}h_{n,1\to k}g_{n,1} \\
\sum_{n=1}^{N}\delta_{n,2}h_{n,2\to k}g_{n,2} \\
\vdots \\
\sum_{n=1}^{N}\delta_{n,M}h_{n,M\to k}g_{n,M}\\
\end{smallmatrix}),
\end{align}
 where $\delta_{n,m}\in \{0,1\}$, $\delta_{n,m}=0$ indicates the $n$-th PA on the $m$-th waveguide is deactivated, and $\delta_{n,m}=1$ denotes PA is activated.

\subsection{Expression of Received Signal}

In RSMA systems, the data streams of users consists of two parts, i.e., private stream and common stream. The private data stream consists of $K$ private messages, and each private message can only be decoded by the corresponding user, while the private messages of other users are treated as interference. The common stream can be decoded by all users\cite{mao2022rate}. In this letter, assume that the transmit power of activated PAs is equally allocated. $\mathbf{L}=[\sqrt{1\text{/}\sum_{n=1}^{N}\delta_{n,1}}, \sqrt{1\text{/}\sum_{n=1}^{N}\delta_{n,2}}, \dots, $ $\sqrt{1\text{/}\sum_{n=1}^{N}\delta_{n,M}}]^T
$ represents the power allocation coefficient for activated PAs on each waveguide. Therefore, the received signal at user $k$ can be expressed as
\begin{align}\label{eq1}
{y}_k=&\underbrace{\mathbf{h}_{k}^H(\mathbf{L}\odot\mathbf{w}_{k,p})s_{k,p}+(\mathbf{L}\odot\mathbf{w}_{c})s_c)}_{\text{desired signal}}\nonumber\\
&+\underbrace{\sum_{i=1,i\neq k}^{K}\mathbf{h}_{k}^H(\mathbf{L}\odot\mathbf{w}_{i,p})s_{i,p}}_{\text{interference signal}}+\underbrace{{n}_k}_{\text{noise}},
\end{align}
where $\mathbf{w}_{k,p}$ and $\mathbf{w}_{c}$ are the private beamforming vector for user $k$ and the common beamforming vector for all users, respectively. $s_{k,p}$ and $s_c$ are the private signal for user $k$ and the common signal with normalized power, respectively. ${n}_k$ is the additive white Gaussian noise with mean 0 and variance $\sigma^2$.

\subsection{SINR Expressions}

For downlink RSMA system, users first decode the common stream and treat all private stream as noise. Hence, the achievable rate for user $k$ decoding the common stream can be expressed as
\begin{align}
R_{k,c}=\log_2(1+\frac{|\mathbf{h}_{k}^{H}(\mathbf{L}\odot\mathbf{w}_{c})|^2}{\sum_{i=1}^{K}|\mathbf{h}_{k}^{H}(\mathbf{L}\odot\mathbf{w}_{i,p})|+\sigma^2_k}).
\end{align}

To ensure that all users can successfully decode the common stream, the common rate should satisfy $R_c=min\{R_{1,c},R_{2,c},...,R_{K,c}\}$.

After the $k$-th user decodes the common stream, the SIC technology is used to remove common stream from the received signal. Therefore, the achievable rate expression for the $k$-th user to decode the corresponding private stream can be expressed as
\begin{align}
R_{k,p}=\log_2(1+\frac{|\mathbf{h}_{k}^{H}(\mathbf{L}\odot\mathbf{w}_{k,p})|^2}{\sum_{i=1,i\neq k}^{K}|\mathbf{h}_{k}^{H}(\mathbf{L}\odot\mathbf{w}_{i,p})|+\sigma^2_k}).
\end{align}

The achievable rate of the $k$-th user is $R_k=r_{k,c}+R_{k,p}$, where $r_{k,c}$ is the common rate of the $k$-th user with $\sum_{k=1}^Kr_{k,c}\leq R_{c}$.

\section{Solution to PA Selection and Waveguide Beamforming}

In order to maximize sum rate, the original problem is divided into two subproblems.  Specifically, a spatial correlation and distance based scheme is leveraged to determine PA activation states, and a SCA based SDP algorithm is proposed to optimize the waveguide beamforming.

\subsection{Problem Formulation}
In this letter, the sum rate problem is optimized, where the power budget, the quality of service, and rate allocation are considered. Then, the optimization problem can be expressed as
\begin{align}
\max_{\mathbf{w}_{c},\mathbf{w}_{k,p},r_{k,c},\delta_{n,m}}: &~ R_c+\sum_{k=1}^{K}R_{k,p}\nonumber\\
s.t. &~R_{k,p}+r_{k,c} \geq R_{min},\label{1req}\\
&~\sum_{k=1}^{K}r_{k,c} \leq R_{c},\label{1com}\\
&~||\mathbf{w}_c||^2+\sum_{k=1}^{K}||\mathbf{w}_{k,p}||^{2}\leq P_{max},\label{1power}\\
&~\delta_{n,m} \in \{0,1\},\label{1PA}\
\end{align}
where (\ref{1req}) ensures the minimum transmission rate requirement $R_{min}$ is met, (\ref{1com}) ensures all users can decode the common rate, (\ref{1power}) constrains the total transmission power budget $P_{max}$, and (\ref{1PA}) indicates whether PAs are activated or not.

\subsection{PA Activation Selection}
In this section, a spatial correlation and distance based scheme is proposed to determine the activation states of PAs. Specifically, initially one PA is activated per waveguide. To minimize path loss, initial PAs are selected based on the total distance to all users. Then, the problem of initial PA selection can be formulated as

\begin{align}
(n^*,m^*)=\mathop{\arg\min}\limits_{n\in\{1,2,\dots,N\}}d_{n,m},  \forall m \in \{1,2\}\label{distance},
\end{align}
 where $(n,m)$ is the index of the $n$-th PA on the $m$-th waveguide, and $d_{n,m}=\sum_{k=1}^{K}|\chi_{n,m}^{PA}-\chi_{k}|$ is the sum of the distances from the $n$-th PA on the $m$-th waveguide to all users.

Then, the PA $(n,m)$ activation status can be updated by
\begin{align}
\delta_{n,m} = \begin{cases}
1 & \text{if } (n,m)= (n^*,m^*),\forall n,m \\
0 & \text{else }
\end{cases}.
\end{align}

At this stage, the spatial correlation can be given by
\begin{align}
\rho=\sum_{k=1}^{K}\sum_{i=k}^{K}\frac{|\mathbf{h}_{k}^{H}\mathbf{h}_{i}|}{||\mathbf{h}_{k}^{H}||||\mathbf{h}_{i}||}\label{spatial}.
\end{align}

The set of activated PAs is denoted as $\psi_{1}$ and the set of candidate PAs is denoted as $\psi_{2}$ , where $\psi_1\cap\psi_2=\Omega$, $\Omega$ is the set of all PAs. After determining the initially activated PAs, remove the PA $(n^*,m^*)$  from $\psi_{2}$ and add it to $\psi_{1}$.

For the remaining PAs, a spatial correlation and distance based method is proposed to determine the states of PAs. Specifically, similarly to (\ref{distance}), select PA $(n^*,m^*)$ in $\psi_2$ with the smallest distance to all users, and remove it from $\psi_2$.  According to (\ref{channel}), the channel state information $\mathbf{h}_{k}^*$ can be updated, and the sum of spatial correlation $\rho^*$ can be obtained by (\ref{spatial}). If $\rho^*$ is less than $\rho$, PA $(n^*,m^*)$ will be added to $\psi_1$. Repeat the above steps until $\psi_2=\emptyset$. The procedure of the proposed method for PA selection is given in Algorithm 1.

\subsection{Waveguide Beamforming}
After the active PAs are determined, the original problem becomes a non-convex problem with respect to the beamforming vector. In order to address this issue, the problem can be transformed to a semidefinite programming  as
\begin{align}
\max_{\mathbf{W}_{c},\mathbf{W}_{{k,p}},r_{k,c}}: &~ R_c+\sum_{k=1}^{K}R_{k,p}\label{wavebeamforming}\\
s.t. &~R_{k,p}+r_{k,c} \geq R_{min},\label{beam1}\\
&~\sum_{k=1}^{K}r_{k,c} \leq R_{c},\label{beam2}\\
&~\text{Tr}(\mathbf{W}_c)+\sum_{k=1}^{K}\text{Tr}(\mathbf{W}_{k,p})\leq P_{max},\label{beam3}\\
&~ \text{rank}(\mathbf{W}_c)= 1,\label{rank1c}\\
&~ \text{rank}(\mathbf{W}_{k,p})= 1,\label{rank1p}\
\end{align}
where $R_{k,c}=\log_2(1+{\frac{\text{Tr}(\mathbf{H}_{k}\mathbf{W}_c)}{\sum_{i=1}^{K}\text{Tr}(\mathbf{H}_{k}\mathbf{W}_{i,p})+\sigma_{k}^{2}}})$, $R_{k,p}=\log_2(1+{\frac{\text{Tr}(\mathbf{H}_{k}\mathbf{W}_{k,p})}{\sum_{i=1,i\neq k}^{K}\text{Tr}(\mathbf{H}_{k}\mathbf{W}_{i,p})+\sigma_{k}^{2}}})$,  $\mathbf{H}_{k}=(\mathbf{L}\odot\mathbf{h}_{k})(\mathbf{L}^T\odot\mathbf{h}_{k}^H)$, $\mathbf{W}_{k,p}=\mathbf{w}_{k,p}\mathbf{w}_{k,p}^H$, and $\mathbf{W}_{c}=\mathbf{w}_{c}\mathbf{w}_{c}^H$.

Since (\ref{beam1}), (\ref{beam2}), and (\ref{beam3}) are non-convex, the relaxation variables $\{A_{k,p},B_{k,p},A_{k,c},B_{k,c}\}$ are introduced to decouple the non-convex expression $R_{k,c}$ and $R_{k,p}$, which can be expressed as
\begin{align}
&A_{k,p} \leq \text{Tr}(\mathbf{H}_{k}\mathbf{W}_{k,p}),\\
&B_{k,p} \geq \sum_{i=1,i\neq k}^{K}\text{Tr}(\mathbf{H}_{k}\mathbf{W}_{i,p})+\sigma_{k}^{2},\\
&\log_2(1+\frac{A_{k,p}}{B_{k,p}})\leq R_{k,p},\label{abp}\\
&A_{k,c} \leq \text{Tr}(\mathbf{H}_{k}\mathbf{W}_c),\\
&B_{k,c} \geq \sum_{i=1}^{K}\text{Tr}(\mathbf{H}_{k}\mathbf{W}_{i,p})+\sigma_{k}^{2},\label{abc}\\
&\log_2(1+\frac{A_{k,c}}{B_{k,c}})\leq R_{k,c}.\
\end{align}

However, (\ref{abp}) and (\ref{abc}) are still non-convex. In order to transform the non-convex constraint into a tractable form, the first-order Taylor approximation is invoked. Then, constraints (\ref{abp}) and (\ref{abc}) can be transformed to
\begin{align}
&\log_2(1+\frac{A_{k,p}^{(t)}}{B_{k,p}^{(t)}})+\ln2(\frac{A_{k,p}-A_{k,p}^{(t)}}{A_{k,p}^{(t)}+B_{k,p}^{(t)}}-\frac{A_{k,p}^{(t)}(B_{k,p}-B_{k,p}^{(t)})}{B_{k,p}^{(t)}(A_{k,p}^{(t)}+B_{k,p}^{(t)})}\nonumber\\
& \triangleq \mu_{k,p}\leq R_{k,p},\\
&\log_2(1+\frac{A_{k,c}^{(t)}}{B_{k,c}^{(t)}})+\ln2(\frac{A_{k,c}-A_{k,c}^{(t)}}{A_{k,c}^{(t)}+B_{k,c}^{(t)}}-\frac{A_{k,c}^{(t)}(B_{k,c}-B_{k,c}^{(t)})}{B_{k,c}^{(t)}(A_{k,c}^{(t)}+B_{k,c}^{(t)})}\nonumber\\
&\triangleq \mu_{k,c}\leq R_{k,c},\
\end{align}
where $A^{(t)}$ and $B^{(t)}$ are the Taylor expansion points in the $t$-th iteration.

Therefore, problem (\ref{wavebeamforming}) is transformed to an SDP problem with rank-one constraint, and it can be expressed as
\begin{align}
\max_{\mathbf{W}_{c},\mathbf{W}_{{k,p}},r_{k,c}}: &~ \mu_c+\sum_{k=1}^{K}\mu_{k,p}\label{p3}\\
s.t. &~\mu_{k,p}+r_{k,c} \geq R_{min},\\
&~\sum_{k=1}^{K}r_{k,c} \leq \mu_{c},\\
&~\text{Tr}(\mathbf{W}_c)+\sum_{k=1}^{K}\text{Tr}(\mathbf{W}_{k,p})\leq P_{max},\\
&~ \mathbf{W}_c\succeq 0, \mathbf{W}_{k,p} \succeq 0,\\
&~ \text{rank}(\mathbf{W}_c)= 1,\label{rank1c}\\
&~ \text{rank}(\mathbf{W}_{k,p })= 1.\label{rank1p}\
\end{align}

 After removing constraints (\ref{rank1c}) and (\ref{rank1p}), the beamforming optimization problem becomes a standard SDP problem, which can be efficiently solved by using the MOSKE solver. Although the two constraints are not involved, the obtained beamforming vector still satisfy the rank one constraints, which will be proved as follows.

 \begin{proof}
 The Lagrangian function for problem  (\ref{p3}) without the rank-one constraint can be expressed as
\begin{align}
\mathcal{L}&=\lambda_{1}(\text{Tr}(\mathbf{W}_c)+\sum_{k=1}^{K}\text{Tr}(\mathbf{W}_{k,p})- P_{max})\nonumber\\
&+\lambda_{2}(A_{k,p}-\text{Tr}(\mathbf{H}_{k}\mathbf{W}_{{k,p}}))\nonumber\\
&+\lambda_{3}(\sum_{i=1}^{K}\text{Tr}(\mathbf{H}_{k}\mathbf{W}_{i,p})+\sigma_{k}^2-B_{k,c})\nonumber\\
&+\text{Tr}(\mathbf{S}_{k}\mathbf{W}_{k,p})+\Lambda,\nonumber\
\end{align}
where $\lambda\geq 0$ and $\mathbf{S}_{k}\succeq 0$ are the Lagrange multipliers, $\Lambda$ is the terms independent of $\mathbf{W}_{k,p}$.  Based on the Karush-Kuhn-Tucker (KKT) conditions, the complementary slackness can be expressed as $\mathbf{W}_{k,p}^{*}\mathbf{S}^{*}_k=0$, and the gradient of Lagrangian function with
 respect to $\mathbf{W}_{k,p}$ can be derived and set to zero:
\begin{align}
\nabla_{\mathbf{W}^*_{k,p}}\mathcal{L}=\lambda_{k}\mathbf{I}_{M}+(\lambda_3-\lambda_2)\mathbf{H}_{k}+\mathbf{S}^*_{k}=0.\nonumber\
\end{align}
Therefore, $\text{rank}(\mathbf{W}^*_{k,p})=\text{rank}(\mathbf{W}_{k,p}^*\mathbf{h}_{k}\mathbf{h}_{k}^{H})\leq \text{rank}(\mathbf{h}_{k}\mathbf{h}_{k}^{H})= 1$ can be obtained. Similarly, $\text{rank}(\mathbf{W}_c)=1$ can also be derived.
\end{proof}

\begin{algorithm}[!tbp]
\caption{{PA activation algorithm based on spatial correlation and distance}}
\begin{algorithmic}[1]
\STATE $\textbf{Input}$ $\psi_{1} = \emptyset$, $\psi_{2} = \Omega$, and $h_{n,m\to k}, \forall n,\forall m$
\STATE $\textbf{Output}$ The set of $\psi_{1}$
        \STATE Adjust $\psi_{1} \gets \psi_{1}\cup\mathop{\arg\min}\limits_{n\in\{1,2,\dots,N\}}d_{n,m}$,  $\forall m$
        \STATE Adjust $\psi_{2}\gets\psi_{2}\ \setminus \mathop{\arg\min}\limits_{n\in\{1,2,\dots,N\}}d_{n,m}$, $\forall m$
        \STATE do $\rho =\sum_{k=1}^{K}\sum_{i=k}^{K}\frac{|\mathbf{h}_{k}^{H}\mathbf{h}_{i}|}{||\mathbf{h}_{k}^{H}||||\mathbf{h}_{i}||}$
        \STATE $\textbf{repeat:}$
        \STATE \quad $(n^*,m^*)=\mathop{\arg\min}\limits_{ (n,m)\in\psi_{2}}d_{n,m}$, $\psi_{2}\gets\psi_{2}\setminus (n^*,m^*)$
        \STATE \quad if $(n,m)=(n^*,m^*)$ or $(n,m)\in\psi_{1}$
        \STATE \quad\quad $\delta_{n,m}=1$
        \STATE \quad else
        \STATE \quad\quad $\delta_{n,m}=0$
        \STATE \quad Update $\mathbf{h}_k^*=( \begin{smallmatrix}
\sum_{n=1}^{N}\delta_{n,1}h_{n,1\to k} g_{n,1} \\
\sum_{n=1}^{N}\delta_{n,2}h_{n,2\to k} g_{n,2} \\
\vdots \\
\sum_{n=1}^{N}\delta_{n,M}h_{n,M\to k} g_{n,M}\\
\end{smallmatrix} )$
        \STATE \quad do $\rho^{*}=\sum_{k=1}^{K}\sum_{i=k}^{K}\frac{|(\mathbf{h}^*_{k})^{H}\mathbf{h}_{i}^*|}{||(\mathbf{h}_{k}^*)^{H}||||\mathbf{h}_{i }^*||}$
        \STATE \quad if $\rho^* < \rho$
        \STATE \quad \quad $\psi_{1}\gets\psi_{1}\cup(n^*,m^*)$
        \STATE $\textbf{until}$~ $\psi_{2}=\emptyset$
        \STATE $\textbf{Output}$ $\psi_{1}$
    \end{algorithmic}
\end{algorithm}

\subsection{Convergence and Complexity Analysis}
 For the PA activation, the proposed algorithm achieves a computational complexity of ${O}_{1}=\mathcal{O}(MN-2)$, significantly lower than the complexity of the exhaustive search approach $\mathcal{O}((2^{N}-1)^M)$. The complexity of the PA beamforming optimization problem is ${O}_{2}=\mathcal{O}({I}_{\max}{\max(N,2(M+1))}^{4}\sqrt{N}log_2 \frac{1}{\epsilon})$, where ${I}_{\max}$ is the iteration number for problem (\ref{p3}) and $\epsilon$ is the accuracy. The total computational complexity of the proposed algorithm is $({O}_1+{O}_2)$.

\section{Simulation Results}
\begin{table}[!tbp]
\caption{Simulation Parameters}\label{simulation}\vspace{-0.3cm}
\centering
\scalebox{0.9}{
\begin{tabular}{c|c}
\hline \hline
Parameters & Values \\
\hline
The height of waveguides $Z^{PA}$ and $Z^{FP}$ & 3m \\
The carrier frequency $f_c$ & 2.8Ghz\\
The number of pinching antennas $N$ & $20 \sim 100$ \\
The length of waveguide $D$ & 20m\\
The location of FP $(X^{FP}_{m},Y^{FP}_{m})$ & (0,-5,3), (0,5,3)m\\
The deployment region of users $(x_{k},y_{k})$ & $(0,-10) \sim (20,10)$m\\
The minimum rate requirement of user $R_{min}$ & 0.1 bps/Hz\\
The transmission power budget $P_{max}$ & $10$ dBm\\
Gaussian noise power $\sigma^2$ & -80 dBm \\
\hline
\end{tabular}}
\vspace{-0.3cm}
\end{table}

In this section, several benchmarking methods are compared to evaluate the rate performance of the proposed spatial correlation and distance based PAs aided RSMA system with multiple waveguides (\textbf{SD-RSMA}).
Fully Activated PAs (\textbf{FAP-RSMA}): All PAs are activated for signal transmission;
Distance based scheme (\textbf{D-RSMA}): A distance based PA activation scheme is employed to assist RSMA system, where only one PA is activated on each waveguide;
PAs assisted NOMA system (\textbf{PA-NOMA}): The PAs aided NOMA system with multiple waveguides;
Conventional Antenna (\textbf{CA-RSMA}): $M$ conventional antennas are deployed at the BS.
 The other simulation parameters are given in Table I.

\begin{figure}[!tbp]
  \centering
  \includegraphics[width=0.35\textwidth]{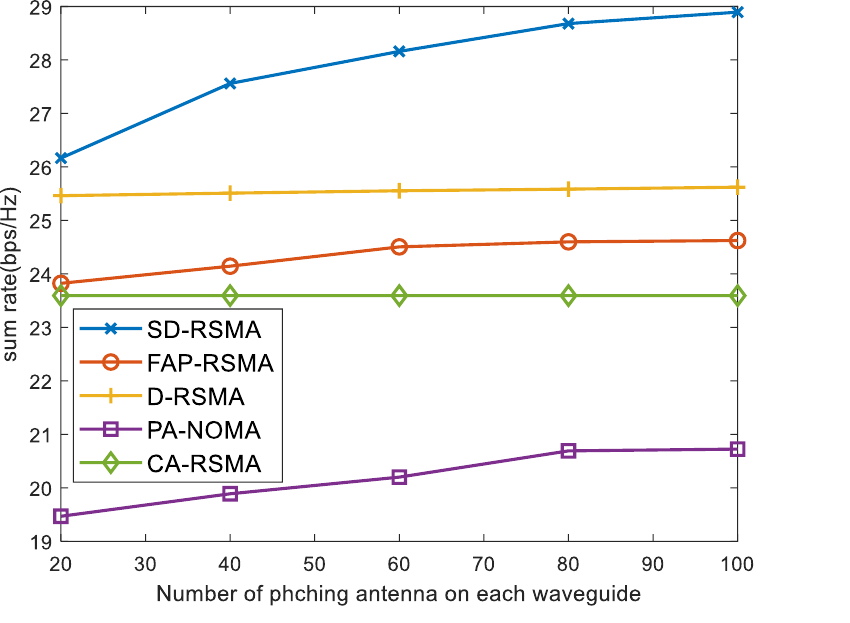}
  \vspace{-0.3cm}
  \caption{The sum rate versus the number of PAs on each waveguide with $D=20$m.}\label{WANG_fig2}\vspace{-0.4cm}
\end{figure}

Fig. \ref{WANG_fig2} depicts the sum rate versus the number of PAs on each waveguide with $D=20$m. Compared with the baselines, the proposed scheme demonstrates superior performance. For FAP, partial PAs located farther from the users are also activated, which results in the higher path loss outweighing the system improvement brought by the degrees of freedom provided by PAs. In contrast to FAP, distance based scheme only consider the distance, while neglecting spatial degrees of freedom. Furthermore, since this scheme determines PA activation only based on distance, the system performance does not improve with increasing number of PAs. Compared to NOMA, RSMA splits the user rate into common rate and private rate, thereby increasing system degrees of freedom. Consequently, RSMA achieves higher sum rate than NOMA owing to these advantages. Compared to conventional antenna systems, the proposed scheme can reduce path loss by deploying PAs closer to user equipments, thereby improving the sum rate performance.

\begin{figure}[!tbp]
  \centering
  \includegraphics[width=0.35\textwidth]{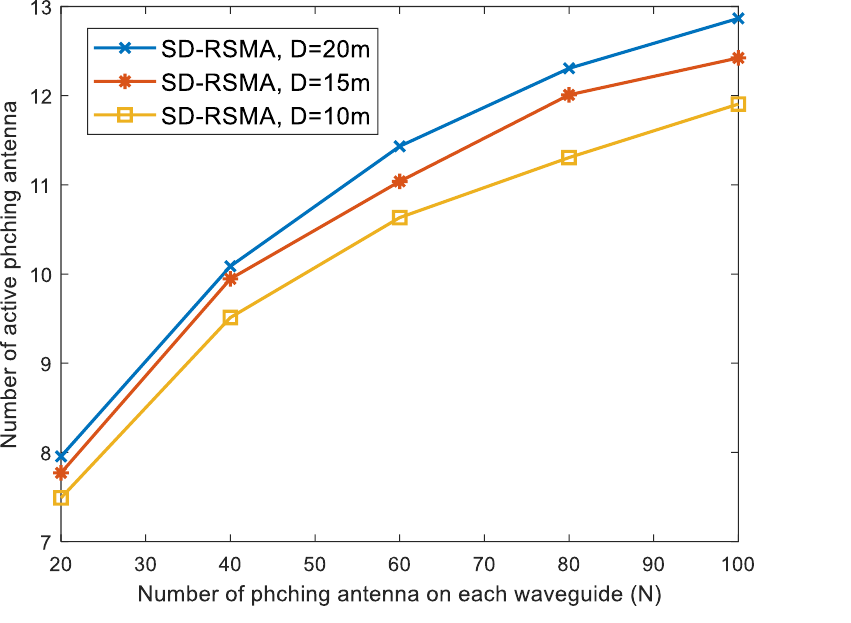}
  \vspace{-0.3cm}
  \caption{The number of active PAs versus the number of PAs on each waveguide.}\label{WANG_fig3}\vspace{-0.4cm}
\end{figure}

Fig. \ref{WANG_fig3} depicts the number of active PAs versus the number of PAs on each waveguide. The increased number of PAs enhances channel selection flexibility, which reduces spatial correlation and allows more PAs to be activated. Conversely, when both waveguide length and user deployment range decrease, the resulting high user density leads to significantly enhanced channel spatial correlation. In this case, the selection of suitable PAs becomes more difficult, resulting in a reduction in the number of active PAs.

\begin{figure}[!tbp]
  \centering
  \includegraphics[width=0.35\textwidth]{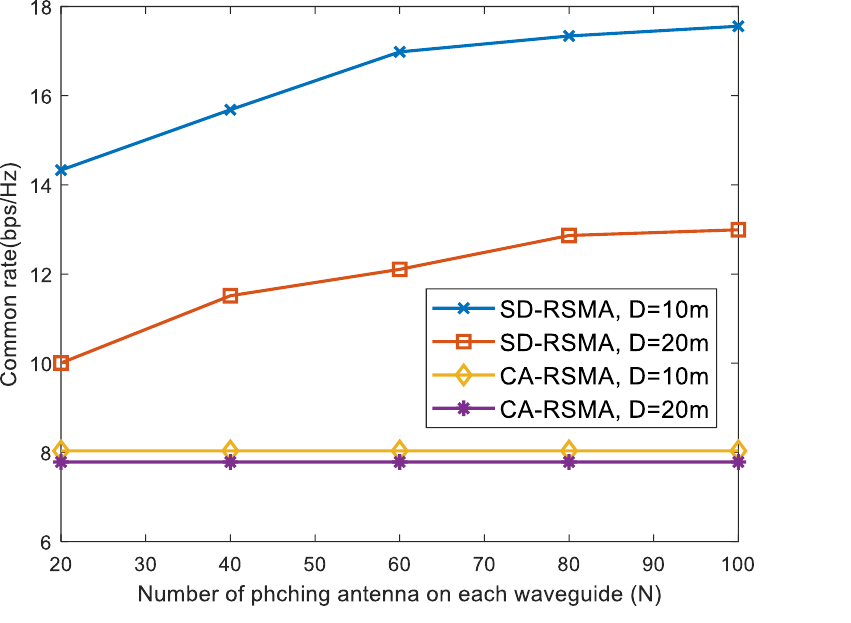}
  \vspace{-0.3cm}
  \caption{The common rate versus the number of PAs on each waveguide.}\label{WANG_fig4}\vspace{-0.4cm}
\end{figure}

Fig. \ref{WANG_fig4} depicts the common rate versus the number of PAs on each waveguide. Unlike conventional antenna-assisted RSMA systems, the PAs assisted RSMA system enhances users' channel quality by activating PAs near the target user to reduce path loss, thereby improving the achievable common rate.

\section{Conclusion}
In this letter, the PAs assisted RSMA system with multiple waveguides is proposed to maximize sum rate. In order to improve the private rate, the spatial correlation is minimized by selecting the PA activation scheme. Meanwhile, the distance from PAs to users is considered to mitigate path loss, thereby enhancing the common message rate. Furthermore, a SDP based SCA algorithm is proposed to optimize the waveguide beamforming. Simulation results demonstrate PAs assisted RSMA system can leverage the deployment flexibility of PAs to enhance performance.

\end{document}